\documentclass[superscriptaddress,longbibliography]{revtex4-2}

\usepackage{graphicx}
\usepackage{dcolumn}
\usepackage{bm}
\usepackage{multirow}
\usepackage{color}

\begin{document}

\title{Spin Dynamics, Loop Formation and Cooperative Reversal in Artificial Quasicrystals with Tailored Exchange Coupling}


\author{Vinayak~Shantaram~Bhat}\email{vbhat@magtop.ifpan.edu.pl}
\affiliation{Laboratory of Nanoscale Magnetic Materials and Magnonics, Institute of Materials (IMX), \'Ecole Polytechnique F\'ed\'erale de Lausanne (EPFL), 1015 Lausanne, Switzerland}
\affiliation{International Research Centre MagTop, Institute of Physics, Polish Academy of Sciences, 02668 Warsaw, Poland}
\author{Sho~Watanabe}
\affiliation{Laboratory of Nanoscale Magnetic Materials and Magnonics, Institute of Materials (IMX), \'Ecole Polytechnique F\'ed\'erale de Lausanne (EPFL), 1015 Lausanne, Switzerland}
\author{Florian~Kronast}
\affiliation{Helmholtz-Zentrum Berlin f\"{u}r Materialien und Energie, Albert-Einstein-Strasse 15, 12489, Berlin, Germany}
\author{Korbinian~Baumgaertl}
\affiliation{Laboratory of Nanoscale Magnetic Materials and Magnonics, Institute of Materials (IMX), \'Ecole Polytechnique F\'ed\'erale de Lausanne (EPFL), 1015 Lausanne, Switzerland}
\author{Dirk Grundler}\email{dirk.grundler@epfl.ch}
\affiliation{Laboratory of Nanoscale Magnetic Materials and Magnonics, Institute of Materials (IMX), \'Ecole Polytechnique F\'ed\'erale de Lausanne (EPFL), 1015 Lausanne, Switzerland}
\affiliation{Institute of Electrical and Micro Engineering (IEM), \'Ecole Polytechnique F\'ed\'erale de Lausanne (EPFL), 1015 Lausanne, Switzerland}
\date{\today}

\begin{abstract}
Aperiodicity and un-conventional rotational symmetries allow quasicrystalline structures to exhibit unprecedented physical and functional properties. In magnetism, artificial ferromagnetic quasicrystals exhibited knee anomalies suggesting reprogrammable magnetic properties via non-stochastic switching. However, the decisive roles of short-range exchange and long-range dipolar interactions have not yet been clarified for optimized reconfigurable functionality. We report broadband spin-wave spectroscopy and X-ray photoemission electron microscopy on different quasicrystal lattices consisting of ferromagnetic Ni\textsubscript{81}Fe\textsubscript{19} nanobars arranged on aperiodic Penrose and Ammann tilings with different exchange and dipolar interactions. We imaged the magnetic states of partially reversed quasicrystals and analyzed their configurations in terms of the charge model, geometrical frustration and the formation of flux-closure loops. Only the exchange-coupled lattices are found to show aperiodicity-specific collective phenomena and non-stochastic switching. Both, exchange and dipolarly coupled quasicrystals show magnonic excitations with narrow linewidths in minor loop measurements. Thereby reconfigurable functionalities in spintronics and magnonics become realistic.
\end{abstract}

\maketitle
%
%

\section*{Introduction}

\begin{figure}
	\includegraphics[width=0.98\textwidth]{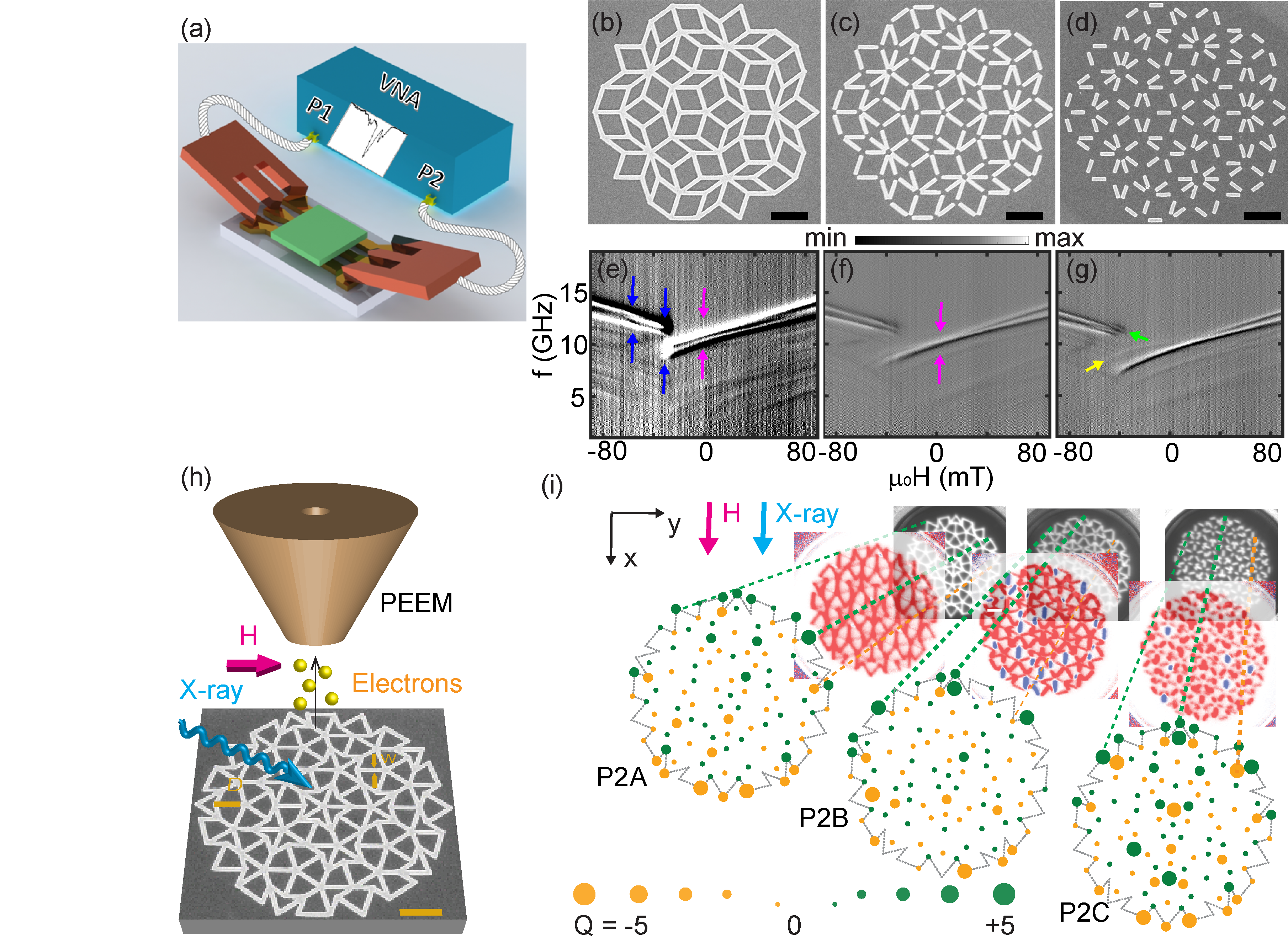}
		\caption{(a) Sketch of broadband spin wave spectroscopy. Scanning electron microscopy (SEM) images showing 3rd generation Penrose P3 lattices: (b) P3A for which nanobars are fully connected, (c) P3B with partially connected nanobars, and (d) P3C with disconnected nanobars. The scale bars correspond to 1~$\mu$m. Gray-scale spin wave spectra obtained on 8th generation (e) P3A, (f) P3B, and (g) P3C samples. The magnetic field was applied along the horizontal direction of graphs (b) to (d) and varied from +90 mT to - 90 mT in a step-wise manner. In the field regime between the blue arrows [Fig. (e)] the reversal of the AMQ takes place. The magenta color arrows mark $H=0$ in (e). The green arrow in (g) marks the high-frequency mode in the reversal regime. The yellow arrow highlights the branch attributed to nanomagnets being almost perpendicular to the applied field. (h) Sketch of the XPEEM imaging experiment performed on ferromagnetic quasicrystals. Here a Penrose P2 tiling is shown as an SEM image. \textit{D} and \textit{w} represent the length and width of a nanobar, respectively. (i) (Back row) PEEM topography images of nanobars arranged on Penrose P2 lattices P2A (left), P2B (center), and P2C (right). (Center row) XPEEM images taken on P2A, P2B, and P2C (from left to right). (Front row) Analysis of the magnetic configurations of vertices based on the charge ($Q$) model. The magnitude of $Q$ is given by diameter of circles (legend). Orange (green) color indicates negative (positive) charge. The broken lines guide the eye for the allocation of charges to a specific vertex. For the configurations shown the maximum evaluated $|Q|$ amounted to 3. The XPEEM images were taken at $\mu_0H=0~$mT after applying $\mu_0H=-52~$mT such that $\mathbf{H}$ had pointed in $-x$-direction. The bright (dark) regions in (h) represent Py (Si substrate). }\label{Fig1a}
\end{figure}

Quasicrystals exhibit aperiodic long-range order and unconventional rotational symmetry, but no translational invariance. Since their discovery \cite{shechtman1984metallic,goldman2013family} the impact of aperiodicity on fundamental physical phenomena is pursued with great interest \cite{goldman2013family}. One powerful avenue to gain novel insight resides in the materials-by-design approach making use of nanofabrication and imaging techniques \cite{bhat2013controlled}. Thereby microscopic understanding of e.g. geometrical frustration in spin ice systems was achieved. The artificial spin ices (ASIs) consisted of either disconnected or interconnected ferromagnetic nanobars that were arranged on strictly periodic lattices with translational invariance \cite{wang2006artificial,qi2008direct,mengotti2011real}. In both types of ASIs cooperative phenomena were found and analyzed via spin ice rules, charge model and energy minimization through clockwise (CW) and counter-clockwise (CCW) flux-closure loops (FCLs) \cite{castelnovo2008magnetic,branford2012emerging}. Vortex-like microstates, i.e., flux-closure loops in a lattice (which are sometimes called microvortices), indicated energy minimization due to magnetic coupling \cite{Keswani2019}. Tailored {\em dipolar} interaction was named key for novel devices \cite{AdvASI2020} based on e.g. reproducible microstates upon cycling an applied magnetic field \cite{gilbert2015PRB,PhysRevB.100.214425}. The spin dynamics in periodic ASIs with reconfigurable magnetic configurations and intentionally introduced magnetic defects have already generated enormous interest \cite{PhysRevB.100.214425,gliga2013spectral,PhysRevB.93.134420,jungfleisch2016dynamic,doi:10.1063/1.4978315,Lendinez_2019,PhysRevB.100.054433}. Still Iacocca {\em et al.} pointed out very recently that dipolar coupling in real ASI might not be sufficient for reconfigurable magnon waveguides \cite{iacocca2020tailoring} consistent with earlier experiments \cite{Li_2016}. For artificial magnetic quasicrystals (AMQs) unconventional magnetic properties and non-stochastic switching were reported for both interconnected lattices with exchange coupling and lattices with edge-to-edge separations of up to about 150 nm between nanomagnets exhibiting dipolar interaction only \cite{bhat2013controlled,brajuskovic2016real,farmer2016direct,bhat2014non,shi2018frustration}. Following the orthodox understanding \cite{WangSci2016} and recently performed micromagnetic simulations one anticipates aperiodicity-induced phenomena for edge-to-edge separations even wider than 150 nm \cite{shi2018frustration}. Particularly, aperiodic quasicrystals promise a plethora of reconfigurable magnetic configurations due to non-stochastic switching in a global magnetic field, while periodic lattices would require the serial writing process based on a magnetic force microscope \cite{WangSci2016}. Domains and domain walls between differently oriented lattice segments of ASIs have been foreseen already as conduits which steer magnons in a reconfigurable manner inside the magnetic lattice \cite{Demo2017} or in an underlayer \cite{iacocca2020tailoring}. However, the following questions are unanswered: (1) How do the cooperative phenomena show up in real samples when one systematically varies the type and relative strength of coupling among aperiodic nanobars? (2) What is the origin of the knee-like anomalies which were reported for the magnetic hysteresis of quasicrystalline Penrose P2 tilings? \cite{bhat2013controlled} (3) How do functional properties of ferromagnetic Penrose P2 and P3 tilings compare? They belong to the same class of ten-fold rotationally symmetric quasicrystal lattices but consist of different geometric prototiles \cite{penrose1979math}. In this article, we report on dynamic and quasi-static investigations based on broadband spectroscopy and magnetic imaging (Fig. \ref{Fig1a}), respectively, performed on ten-fold rotationally symmetric Penrose P2, P3 and the eight-fold rotationally symmetric Ammann tilings. We address the open questions by magnetic resonance spectra taken in the major loop and the hysteretic regime as well as X-ray photo-electron emission microscopy (XPEEM) using x-ray circular dichroism (XMCD) with high spatial resolution (Methods).\\ \indent  For our studies we prepared artificial magnetic quasicrystals  in the form of nanobars (Fig. \ref{Fig1a}) out of the magnetically isotropic alloy Ni\textsubscript{81}Fe\textsubscript{19} (Py) on Penrose P3 [Fig. \ref{Fig1a}(b) to (d)], P2 [Fig. \ref{Fig1a}(h)], and Ammann (Fig. \ref{Fig1b}) quasicrystal lattices using nanofabrication techniques.
The width $w$ of nanobars, their thickness and intervertex spacing were kept at 120 nm, 25 nm, and 810 nm, respectively. We varied the lengths $D$ of nanobars from sample to sample between 810 nm and 408 nm. Thereby we created quasicrystals which consisted of interconnected (A), partially connected (B) and fully separated nanobars (C), respectively [compare Fig. \ref{Fig1a}(b) to \ref{Fig1a}(d)]. The interconnected nanobars joining in the vertices of an AMQ of kind A were both exchange and dipolarly coupled. Separated nanobars in kind B and C were dipolarly coupled only. In sample B (sample C) opposing nanobars exhibited edge-to-edge separations of up to about 200 nm (400 nm). The imaging presented here show that for non-stochastic switching and cooperative reversal in Penrose and Ammann tilings the recently explored dipolar coupling \cite{shi2018frustration} is not sufficient. We observe significant domain formation only when quasicrystalline Penrose P3 and Ammann tilings are exchange-coupled. Both exchange- and dipolarly coupled lattices show reprogrammable magnonic excitations of narrow linewidth. Our findings are key when designing quasicrystals for field-controlled functionalities exploiting return-point memory and reproducible magnetic states \cite{gilbert2015PRB,Demo2017,Sethna1997}.

\section*{Results and Discussion}
\subsection*{Broadband spectroscopy in the major loop}
For the presentation of the results we decompose the Penrose lattices reported in Fig. \ref{Fig1a} into three types of nanobars, i.e., Type I, Type II, and Type III based on the angle $\phi$ that the nanobars take with respect to the direction of magnetic field \textit{H} used in Fig. \ref{Fig1a}: Type I exhibits $\phi$ = 0\textsuperscript{o}, Type II $\phi$ = $\pm $ 36\textsuperscript{o}, and Type III  $\phi$  = $\pm $ 72\textsuperscript{o}. In the Ammann tilings there exist nanobars with $\phi$ = 90\textsuperscript{o}. Considering Ref.~\cite{aharoni1998demagnetizing} we calculated shape anisotropy fields $\mu_0$\textit{H}\textsubscript{ani} for individual nanobars of lengths 810 nm, 609 nm, and 408 nm and obtained 147 mT, 140 mT, and 117 mT, respectively.
Note that the maximum field $ \mu_{0}$$ |$\textit{H}\textsubscript{max}$| $ = 90 mT that was available in the broadband spin-wave spectroscopy setup was smaller than the calculated fields $\mu_0$\textit{H}\textsubscript{ani}. As a consequence, nanobars perpendicular to the applied field could not be saturated. The estimated nucleation field for incoherent reversal in an ideal isolated nanobar with $D=810~$nm via curling amounted to 70 mT \cite{burn2015angular}. The minimum reversal field for coherent rotation was about the same value. Considering these values, we expected the field regime ranging from $+90$~mT$\leq\mu_{0}H\leq-90~$mT to be large enough to reverse the magnetization of Type I and II nanobars but not of individual nanobars of Type III exhibiting an angle $|\phi|$ of 72\textsuperscript{o}.

\begin{figure}
	\includegraphics[width=0.5\textwidth]{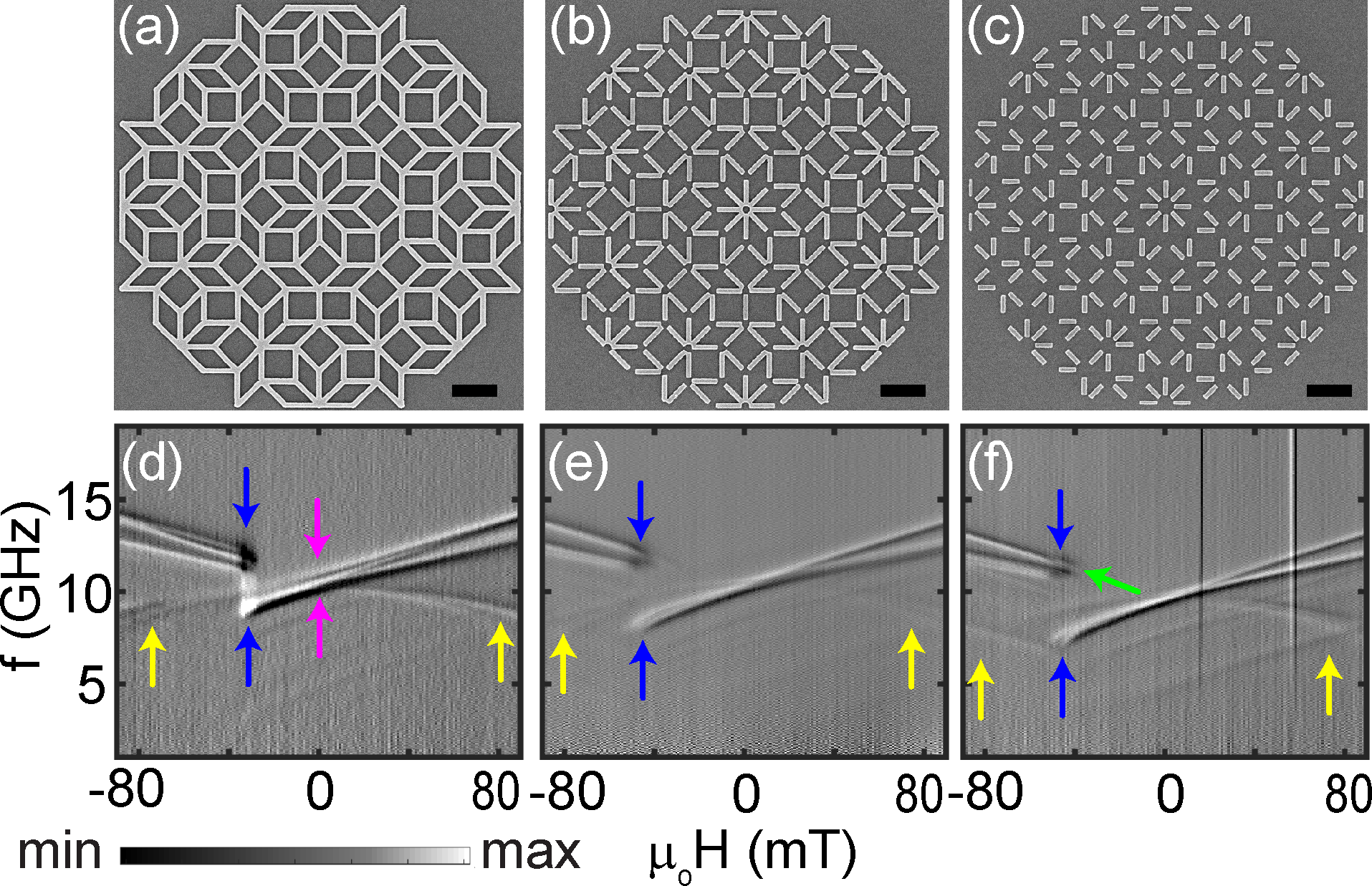}
		\caption{SEM images showing 1st generation Ammann tilings (a) ATA for which nanobars are fully connected, (b) ATB with partially connected nanobars and (c) ATC with disconnected nanobars. The black color scale bar represents 1 micrometer. Gray-scale spin wave spectra obtained on 4th generation (d) ATA, (e) ATB and (f) ATC. The magnetic field was applied along the horizontal direction  of graphs (a) to (c) and varied from +90 mT to - 90 mT in a step-wise manner. In a relatively small field regime near the blue arrow in (d) to (f) the reversal of AT takes place. The magenta color arrows mark $H=0$. The yellow arrows in (d) to (f) highlight the branches attributed to nanomagnets being perpendicular to the applied field. The green arrow in (f) marks the high-frequency mode in the reversal regime of ATC.}\label{Fig1b}
\end{figure}

In the following we present and discuss broadband spectroscopy data obtained on P3 lattices and Ammann tilings for which we have observed large domain formation in the XPEEM imaging experiments. In Fig.~\ref{Fig1a}(e) to (g) we display spectra taken on Penrose P3 tilings when varying the applied magnetic field from +90 mT to -90 mT. For all three samples we see two strong branches at large absolute field values consistent with the interconnected AMQs reported earlier in Ref.~\cite{Bhat2018}. We attribute the branches to resonances in Type I (highest frequency) and Type II (second highest frequency) nanobars. Considering a field of 90 mT, the branch frequencies decrease from (e) to (g), i.e., for the interconnected nanobars (P3A) the two prominent branches reside at overall larger frequencies than for the disconnected nanobars (P3C). From sample P3A to P3C the lengths of nanobars reduce and the demagnetization effect enhances, thereby reducing the internal fields and consequently the resonance frequencies \cite{Gurevich96}. Below the prominent branches manifolds of further resonances are found which are attributed partly to standing spin waves confined along the nanobars. Due to the corresponding backward volume magnetostatic spin wave configuration their resonance frequencies reside at small values. We consider the field regime between the blue arrows in Fig.~\ref{Fig1a}(e) to be the regime in which nanobars of sample P3A reverse. The switching field regime extends from about -30 mT to -65 mT, very similar to a nominally identical interconnected AMQ investigated in Ref. \cite{Bhat2018}. Before we report imaging of magnetic configurations in the reversal regime it is instructive to discuss further details of the spin dynamics in the different quasicrystals.\\
\indent In Fig.~\ref{Fig1a}(e) and (f) the magenta arrows highlight the branches of Type I and Type II nanobars at zero field. In Fig.~\ref{Fig1a}(e) the two branches of P3A are clearly split at $H=0$. This is not the case in P3B. Here the two branches are degenerate at $H=0$, indicating that Type I and Type II experience the same internal magnetic field. The same degeneracy is observed for P3C at $H=0$ in Fig.~\ref{Fig1a}(g). We attribute the frequency difference observed in Fig.~\ref{Fig1a}(e) to the coupling between spin-wave modes in Type I and Type II nanobars leading to an avoided crossing. Nanobars in the interconnected AMQ P3A hence interact. The yellow arrow in Fig.~\ref{Fig1a}(g) highlights a faint branch which exhibits an agility $df/dH>0$ for $H<0$ and, at $H=0$, approaches the degenerate frequencies of Type I and Type II nanobars. The characteristics of this branch are consistent with the field-dependent resonance frequency of Type III nanobars which are at $\pm 72$~deg. For them, $H$ is applied almost along their hard-axis direction. A similar faint branch is seen for P3B in Fig.~\ref{Fig1a}(f). In P3C we resolve a specific high-frequency mode in the reversal regime (green arrow) which is not observed in P3A and P3B and will be discussed after the presentation of the magnetic imaging.\\
\indent In Fig.~\ref{Fig1b}(d) to (f) we display spectra of interconnected (ATA), partially connected (ATB) and disconnected nanobars (ATC). For the Ammann tilings, nanobars of Type I, II and III make angles $ \phi $ = 0\textsuperscript{o}, $ \pm $ 45\textsuperscript{o}, and $ \pm $  90\textsuperscript{o}, respectively. The faint branches highlighted by yellow arrows in Fig.~\ref{Fig1b}(d), (e) and (f) are found in all three Amman tilings ATA, ATB and ATC, respectively. These branches originate from the Type III nanobars which are at an angle of 90 deg with respect to the applied field $H$. $H$ thus points along their hard axis. Again at large absolute fields, two prominent branches are found which are consistent with Type I (high frequency) and Type II (second highest frequency) nanobars. Only for the interconnected AMQ ATA we observe a frequency splitting between Type I and Type II nanobars (magenta arrows) near $H=0$. For AMQs ATB and ATC (disconnected nanobars) the frequency degeneracy occurs. The reversal field regime for ATA [blue arrows in Fig.~\ref{Fig1b}(d)] is found to be narrow compared to P3A [Fig.~\ref{Fig1a}(e)]. The interconnected nanobars reverse close to -30 mT without a significant switching field distribution. The distribution widens for the other (partly) disconnected AMQs. A detailed analysis of spectra of ATC provides again a specific high-frequency mode [green arrow in Fig.~\ref{Fig1b}(f)] in the reversal regime, similar to P3C.
Spin-wave spectra obtained in minor loops and the reconfigurable characteristics of AMQs are discussed after presenting XPEEM experiments by which we image magnetic states in the hysteretic regime.

\subsection*{Magnetic imaging of partly reversed quasicrystals}
\begin{figure}
	\includegraphics[width=0.98\textwidth]{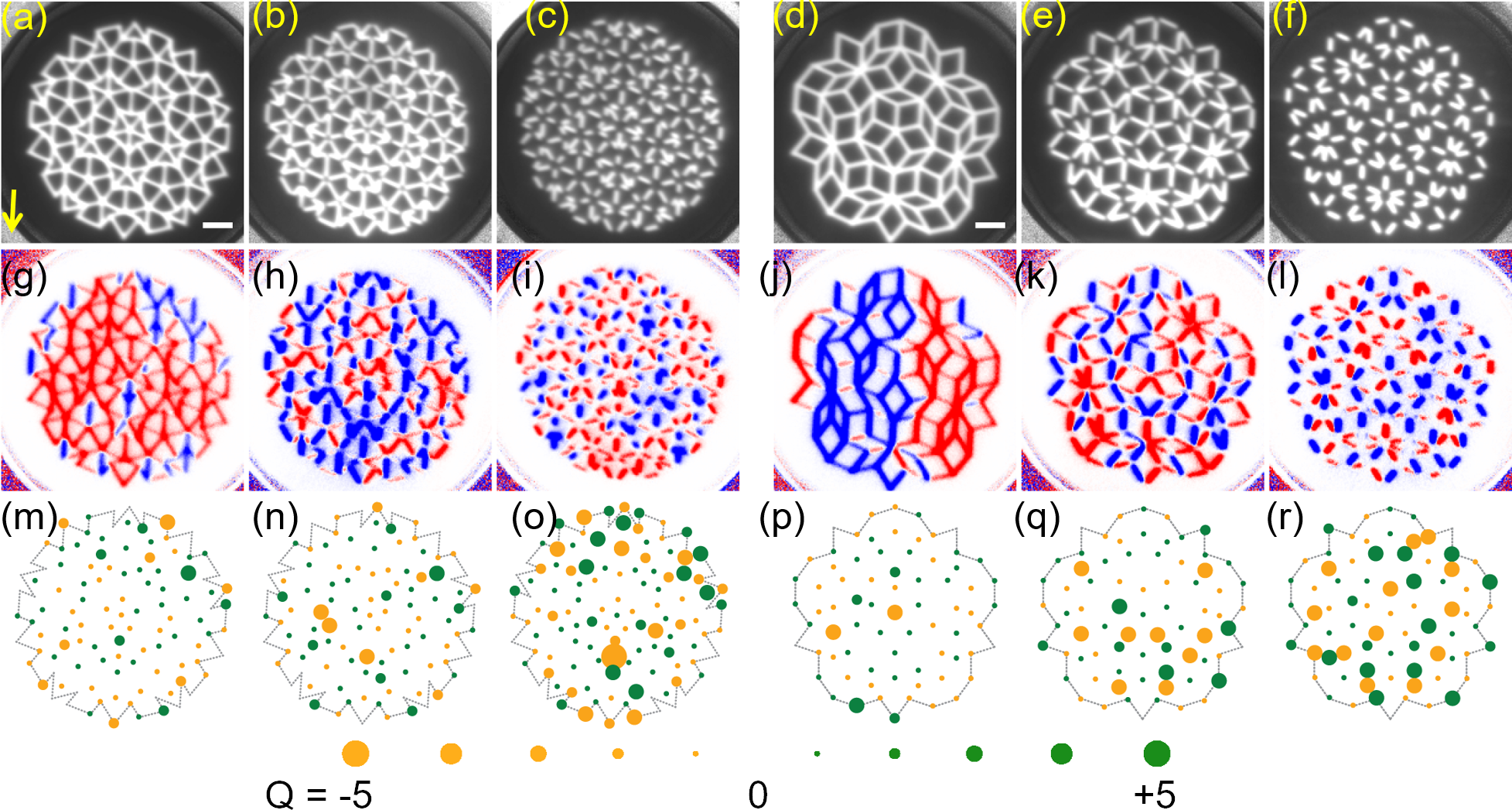}
		\caption{ (a) - (f) XPEEM topography images for six different quasicrystals as labelled in the graphs. Bright (dark) regions correspond to magnetic (non-magnetic) material. The scale bar given in (a) corresponds to 1 $ \mu $m. The arrow in (a) represents the X-ray direction and the magnetic field direction. Selected magnetic images of remnant states using XPEEM after applying different field values $ \mu_{o} H $ (given in parenthesis) for (g) P2A (32.5 mT), (h) P2B (41.6 mT), (i) P2C (42.64 mT), (j) P3A (36.4 mT), (k) P3B (42.9 mT), and (l) P3C (48.1 mT) representing the switching of 44\%, 51\%, 46\%, 56\%, 44\%, and 55\% type I nanobars, respectively. Blue (red) colors represent magnetization parallel (opposite) to the X-ray direction. Blue color indicates a reversed nanobar. The images represent the states attained when about 50\% of Type I nanobars switched. Notice the weak contrast in case of nanobars which do not point along the X-ray direction. (m) – (r) CM analysis of the XMCD experimental data shown in (g) – (l). The green and orange filled circles at the vertices represent positive and negative vertex charges, respectively. Here the circles with smallest and largest diameter represent charge  $|Q|$ = 0 and $|Q|$ = 5, respectively. When quasicrystals consist of disconnected nanobars a large charge of up to $|Q|$ = 5 can be found. The magnetic configurations of P2 and P3 tilings in the as-grown state are displayed in Supplementary Fig. 1.}
\label{Fig2a}
\end{figure}
\begin{figure}
	\includegraphics[width=0.98\textwidth]{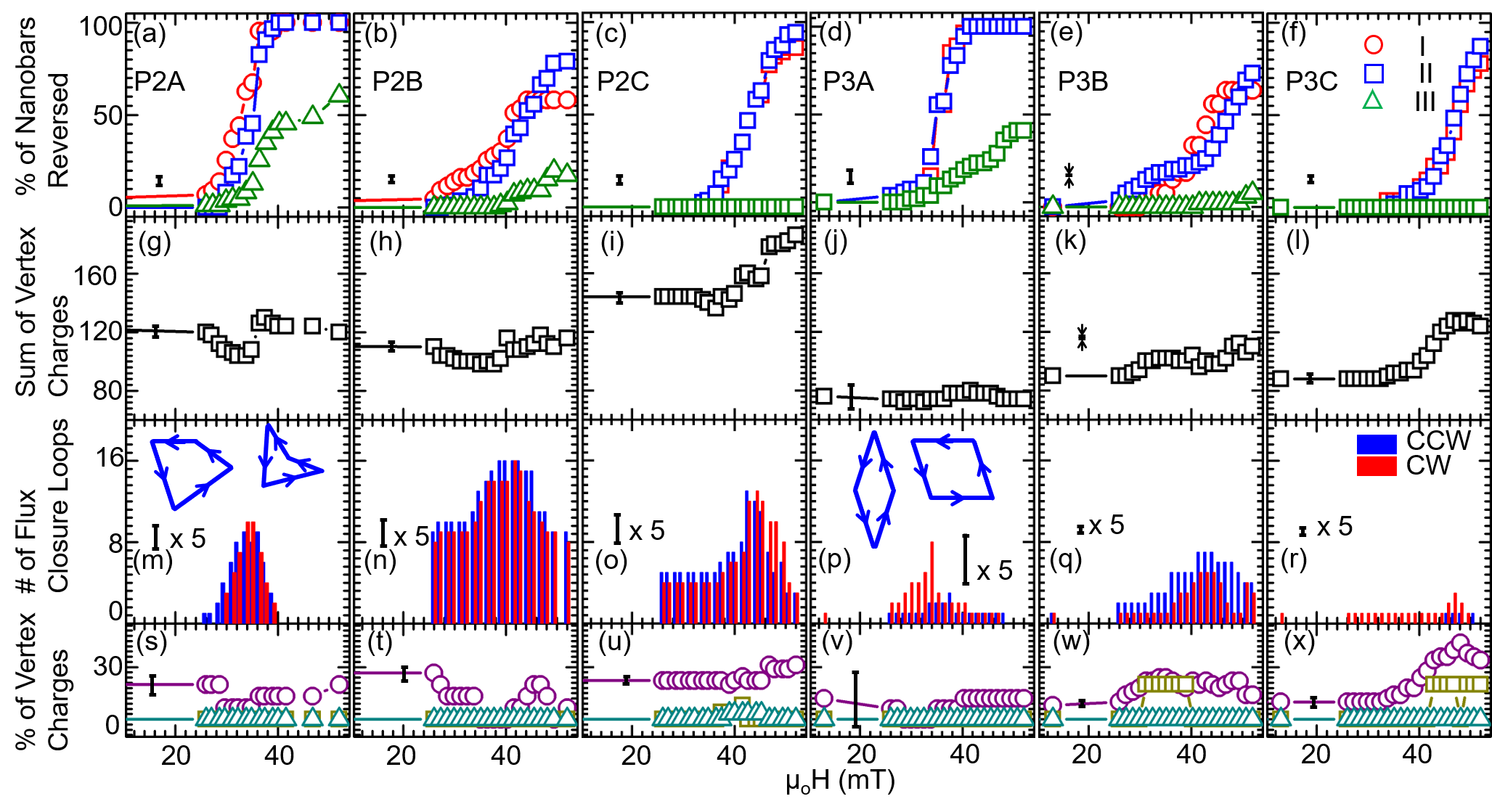}
		\caption{(a) - (f) Field-dependent reversal of Type I (red circle), Type II (blue square) and Type III (green triangle) in AMQs P2A, P2B, P2C, P3A, P3B, and P3C, respectively. (g) - (l) Total charge  \textit{Q\textsubscript{tot}} for P2A, P2B, P2C, P3A, P3B, and P3C, respectively.  (m)-(r) Number of FCLs [see the insets in (m) and (p) for their definition] for P2A, P2B, P2C, P3A, P3B, and P3C, respectively. (s) - (x) Evaluation of relative amounts of maximum possible individual vertex charges \textit{Q} in P2A, P2B, P2C, P3A, P3B, and P3C, respectively. The purple, dark yellow, and dark cyan colored symbols represent charges 3 (for \textit{N} = 3), 4 (for \textit{N} = 4), and 5 (for \textit{N} = 5), respectively. The values in (s) - (w) were multiplied by 3 for better visualization using a unique y scale for all graphs. Notice the presence of charges $|Q|$ = 5 in P2C for \textit{N} = 5 vertices.}\label{Fig3}
\end{figure}
\begin{figure}
	\includegraphics[width=0.98\textwidth]{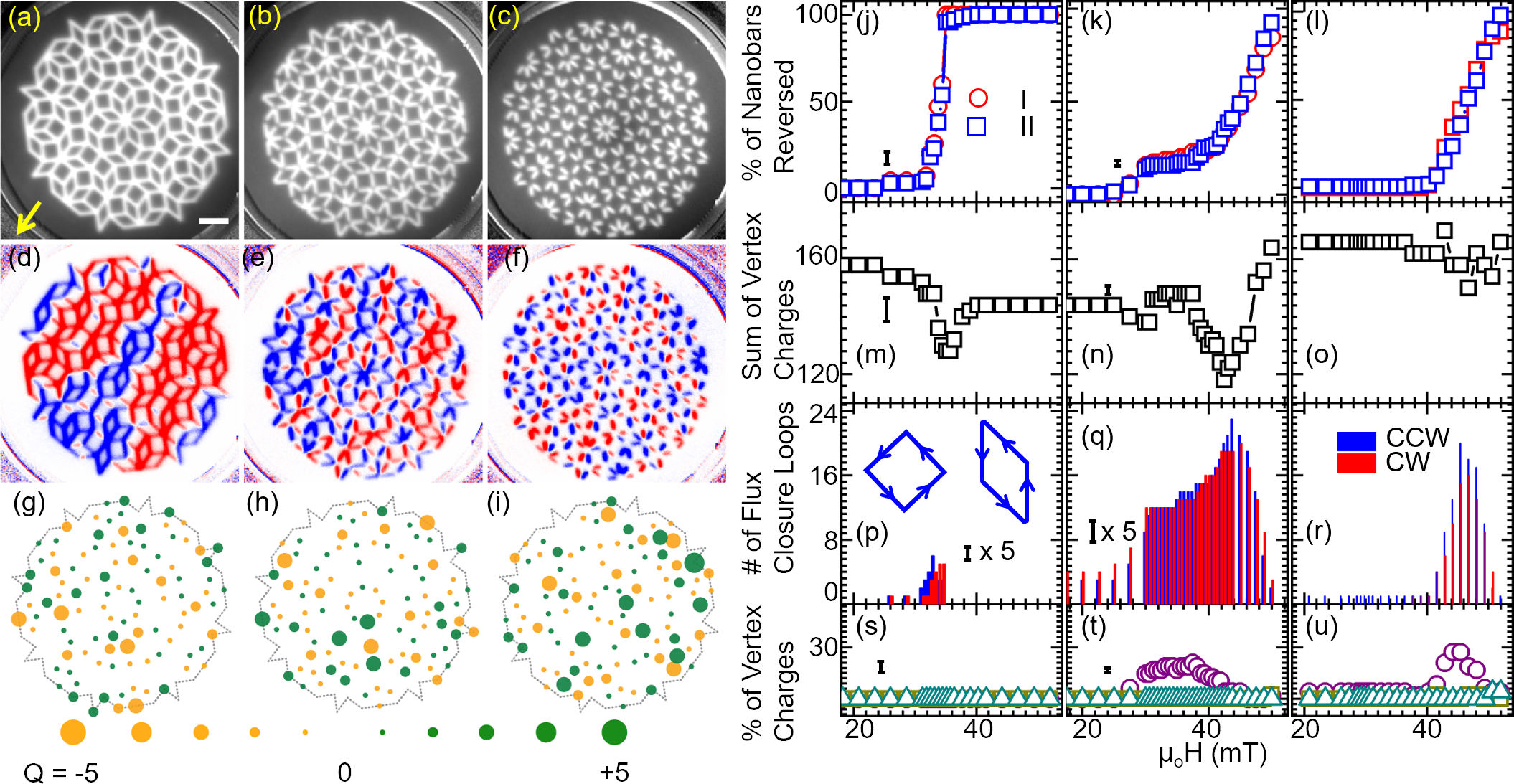}
		\caption{XMCD-PEEM topography image for Ammann tilings (a) ATA ($ \mu_{0} H$ = 34 mT), (b) ATB (at 47 mT), and (c) ATC (at 47 mT) representing the switching of 47\%, 47\%, and 53\% Type I nanobars. Bright (dark) regions correspond to magnetic (non-magnetic) regions. The scale bar corresponds to  1 $ \mu $m. The arrow represents the field direction. The X-ray direction was misaligned by 22\textsuperscript{o} to optimize the detection of nanobars of all the different orientations. (d) - (f) XMCD data of topography images shown in (a) – (f). Blue (red) colors represent magnetization parallel (opposite) to the X-ray direction. Blue color indicates a reversed nanobar. Notice the weak contrast of nanobars whose orientation deviates from the X-ray direction. (g) – (i) CM analysis of XMCD experimental data shown in (d) – (f). The green and orange filled circles at the vertices represent positive and negative vertex charges, respectively. Here the circles with smallest and largest diameter represent charge $|Q|$ = 0 and $|Q|$  = 5, respectively. (j) - (l) Classification of reversal in terms of switching of Type I and II of nanobars for ATA, ATB, and ATC, respectively. Legends I and II represent nanobars of Types I and II.  (m) - (o) Total charge count for ATA, ATB, and ATC, respectively.  (p) - (r) Number of FCLs [see the inset in (p) for their definition] for ATA, ATB, and ATC, respectively. (s) - (u) Charge depiction for ATA, ATB, and ATC, respectively. The purple, dark yellow, and dark cyan colored symbols represent charges 3 (\textit{N} = 3), 4 (\textit{N} = 4), and 5 (\textit{N} = 5), respectively.	}\label{Fig4}
\end{figure}

In the following we discuss magnetic imaging [Fig.~\ref{Fig1a}(h)] of partly reversed AMQs. We have studied the three designs of AMQs introduced previously~\cite{Bhat2018}, i.e., Penrose P2 and P3 lattices as well as Ammann tilings. In the XPEEM microscope the maximum field $ \mu_{0}$$ |$\textit{H}\textsubscript{max}$| $ that was available to magnetize the samples in initial configurations [Fig.~\ref{Fig1a}(i) of P2A and Supplementary Fig. 3] amounted to 52 mT. This value was smaller than the calculated fields $\mu_0$\textit{H}\textsubscript{ani} but larger than the fields that initiated reversal for Type I and Type II nanobars in the broadband spectroscopy experiments of Figs.~\ref{Fig1a} and ~\ref{Fig1b}. Still the field was not large enough to reverse the magnetization of an individual Type III nanobar ($\phi$ = $\pm $72\textsuperscript{o} (Penrose) or 90\textsuperscript{o} (Ammann)) if it did not interact with nanobars of Type I and II.\\ \indent In order to evaluate magnetic states in the saturated and partly reversed quasicrystals, we considered the shape-anisotropy induced bistability (Ising nature) of nanobars and exploited the so-called charge model (CM) (see Methods). We first present experimental data obtained on the Penrose lattices [Fig. \ref{Fig2a}(a) to \ref{Fig2a}(f)]. In Fig. \ref{Fig2a}(g) to \ref{Fig2a}(l) we show XPEEM images taken on magnetic states of P2 and P3 AMQs after initiating partial reversal in a minor loop (as described in Methods). Each image displays a remnant state in zero field after application of a specific field. Images of P2A in Fig. \ref{Fig2a}(g) and P3A in Fig. \ref{Fig2a}(j) show that these AMQs with interconnected nanobars contain domains or chains of reversed nanobars (blue) next to domains in which nanobars remained in the original orientation (red). In particular P3A incorporates a large-area domain of reversed nanobars. The reversed domains and chains are found to include reversed nanobars of Type III though their anisotropy field was estimated to be larger than the applied reversal field. The detection of reversed Type III nanobars reflects a {\em cooperative} phenomenon in the magnetic hysteresis of a quasicrystal. Their reversal is not triggered by the relatively weak external field alone. The reversal is attributed to the influence of neighboring Type I and Type II nanobars which exhibit $ |\phi| $  $\leq $ 36\textsuperscript{o}.
The analysis in terms of the CM (see Supplementary Fig. 4) displayed in Fig. \ref{Fig2a}(m) and \ref{Fig2a}(p) shows that vertices formed by the interconnected nanobars exhibit overall small values of \textit{Q} in their remnant states [the maximum value amounts to \textit{Q} = 3 in Fig. \ref{Fig2a}(m) and \ref{Fig2a}(p)]. \\ \indent AMQs from partially and fully separated nanobars, imaged as P2B [Fig. \ref{Fig2a}(h)], P2C [Fig. \ref{Fig2a}(i)], P3B [Fig. \ref{Fig2a}(k)], and P3C [Fig. \ref{Fig2a}(l)], show numerous reversed magnets that are (more) randomly distributed over the whole AMQ compared to the two AMQs with completely interconnected nanobars. From XPEEM images taken for many different magnetic histories (not shown) we found that switching events in the form of avalanches were seen in the interconnected (exchange-coupled) P2A and P3A samples but were not seen in the disconnected P2C and P3C samples. In the latter samples, opposing nanobars formed fragmented domains which might not be functional for the proposed magnon steering.\\ \indent
In Fig. \ref{Fig3} we analyze the reversal in the AMQs in detail.
In Fig. \ref{Fig3}(a) to \ref{Fig3}(f) the relative numbers of reversed nanobars are depicted as extracted from a series of XPEEM microscopy images taken at remnance after applying different magnetic fields \textit{H}. For P2A and P2B we find reversals to start from Type I nanobars [red symbols in Fig. \ref{Fig3}(a)]. For the other AMQs P2C, P3A, and P3B, Type II nanobars switch first. Once Type I or Type II nanobars reverse in the interconnected P2A or P3A, the reversal of their Type III nanobars follows. A less pronounced successive reversal of Type III nanobars is found for P2B and P3B. Here, the number of reversed Type III nanobars is much smaller compared to P2A and P3A. In P2C and P3C consisting of completely disconnected nanobars we do not find the reversal of Type III nanobars in the accessible field regime. Here, the interaction between nanomagnets is too small.\\ \indent In Fig. \ref{Fig3}(g) to \ref{Fig3}(l) we show the total charge \textit{Q\textsubscript{tot}} extracted from XPEEM images. In AMQs P2A, P2B and P2C the onset of reversal is accompanied by a small global minimum in \textit{Q\textsubscript{tot}}. In case of P2A and P2B \textit{Q\textsubscript{tot}} regains a value close to the initial state at large \textit{H}. This is different for P2C for which \textit{Q\textsubscript{tot}} grows with increasing reversal field. For the Penrose P3 tiling consisting of interconnected nanobars (P3A), \textit{Q\textsubscript{tot}} stays small at large $\mu_0$\textit{H} (vertex charges analyzed for +52 mT are displayed in Supplementary Fig. 4). A growth of \textit{Q\textsubscript{tot}} is observed for P3B and P3C. We attribute the growth in \textit{Q\textsubscript{tot}} to the fact that disconnected nanobars of Type III do not experience cooperative reversal.\\ \indent
In Fig. \ref{Fig3}(m) to \ref{Fig3}(r) we summarize the number of flux-closure loops (FCLs) present in the XPEEM images. For Penrose P2 tilings the minimum in \textit{Q\textsubscript{tot}} is accompanied by a maximum of FCLs. The maximum in FCLs roughly occurs when about 50~\% of Type I and Type II nanobars have undergone switching. For P3B a pronounced maximum in FCLs is found as well. For P3A and P3C the corresponding variation is small. Note that P2B, P2C and P3B exhibit a large number of FCLs already before pronounced switching has taken place. We attribute this observation to the limited field strength (52 mT) that was available to define the initial magnetic states (Supplementary Fig. 3). The maximum field was not large enough to saturate Type III nanobars that were not connected to neighboring nanobars.\\ \indent
In Fig. \ref{Fig3}(s) to \ref{Fig3}(x) we depict the number of individual vertices that exhibit the maximum charge \textit{Q} = 3, \textit{Q} = 4 and \textit{Q} = 5 for neighbor numbers \textit{N} = 3, \textit{N} = 4 and \textit{N} = 5, respectively. For interconnected lattices P2A and P2B the number of vertices with \textit{Q} = 4 and \textit{Q} = 5 is negligible. Strikingly, the number of vertices with \textit{Q} = 3 takes its lowest value when the maximum number of FCLs is reached in P2A. A similar behavior is observed for P2B and P3A. P2C, P3B, and P3C behave differently in that the number of vertices with \textit{Q} = 3 does not go through a global minimum as a function of reversal field. In case of P2C and P3C the number grows with \textit{H}. A considerable number of vertices with \textit{Q} = 4 (\textit{Q} = 5) is found only in P3B and P3C (P2C). Large individual vertex charges hence occur in Penrose P2 and P3 tilings with disconnected nanobars, but not for P2 and P3 tilings with interconnected ones. In the latter cases, vertex configurations with large \textit{Q} [as stabilized in micromagnetic simulations for interconnected nanobars in Fig. 6(a) of the supplementary information] have thus not been observed in the reversal regime. The real AMQ lattices made from interconnected nanobars avoided these high-energy configurations and formed FCLs instead. We argue that the exchange interaction in the vertices of the interconnected P2 and P3 tilings plays the major role for the observed cooperative magnetization reversal which included Type III nanobars.\\ \indent Precursors for the cooperative reversal are the Type I and Type II nanobars that meet at a vertex exhibiting a relatively large charge (e.g. \textit{Q} = 2 for \textit{N} = 4). The occurrence of large \textit{Q} indicates the violation of the local ice rule. Figure \ref{Fig3}(s) to \ref{Fig3}(x) reveals that ice rule violations are pronounced in P2C, P3B and P3C with disconnected nanobars. Our data suggest that weakly interacting nanobars in AMQs provoke ice rule violations. Note that P3B and P3C did not show the frequency splitting between Type I and Type II nanobar resonances near $H=0$ [Fig.~\ref{Fig1a}(f) and (g)]. The absence of splitting is consistent with weakly interacting nanobars. \\ \indent To gain further insight into the violation of spin ice rules we have simulated low- and high-energy vertex states for connected and disconnected nanobars [Supplementary Figs. 6(b) to 6(i)]. We see that for interconnected nanobars found e.g. in P2A, the total vertex energy increases by about 80  \%  from Supplementary Fig. 6(b) (\textit{Q} = -1) to Fig. 6(c) (\textit{Q} = -5), respectively, considering a vertex with \textit{N} = 5. For the disconnected nanobars relevant e.g. in P2C, a violation of the spin ice rule [\textit{Q} = -5 in Supplementary Fig. 6(e)] causes an energy higher by only 7 \% compared to Fig. 6(d) (\textit{Q} = -1). Consistent energy variations with \textit{Q} are found for vertices with \textit{N} = 4 in Supplementary  Fig. 6(f) to 6(i). The large energy cost for a spin ice rule violation found in simulations for fully interconnected nanobars favors low vertex charges which agree with the spin ice rule. Indeed we experimentally detected mainly low-\textit{Q} states in the interconnected lattices of P2A, P3A and ATA in the reversal regime. If present, high-\textit{Q} (high energy) vertices in interconnected AMQs act as nucleation sites for the cooperative reversal. They give rise to non-stochastic switching \cite{bhat2014non}, presumably causing the staircase-like jumps or knee anomalies \cite{bhat2013controlled}. Vertices containing disconnected nanobars are found to accommodate particularly large \textit{Q} and violate the spin ice rule. However these charges do not trigger cooperative reversal which we attribute to the relatively weak dipolar coupling in our AMQs. Still our results indicate that dipolar interaction is enough to provoke energy minimization on a local scale between vertices via flux-closure loops. We find an increase in the number of flux-closure loops in almost any of the investigated quasicrystals in the reversal regime. The dipolar interaction and loop formation are however not sufficient for creating extended domains.\\ \indent
In Fig. \ref{Fig4} we summarize the experimental data obtained on Amman tilings (AT) depicted in Fig. \ref{Fig4}(a), (b) and (c). The initial magnetic configurations are displayed in Supplementary Fig. 5.
The XPEEM experiments are shown in Fig. \ref{Fig4}(d) to \ref{Fig4}(f). For interconnected nanobars we detect a large central domain of reversed nanobars. For the partially and fully disconnected nanobars the reversed nanobars are more distributed over the AMQs. In analogy to disconnected Penrose tilings the analysis based on the CM provides a tendency towards slightly larger individual vertex charges when going from ATA to ATC. Still this tendency is less significant in the graphs of Fig. 5(g) to 5(i) compared to the Penrose tilings. Analyzing all our XPEEM datasets we find that almost all of Type III nanobars with $ \phi $  = $\pm $ 90\textsuperscript{o} do not undergo switching in Ammann tilings (compare also data taken at +52 mT and shown in Supplementary Fig. 5). Studying the reversal of Type I and Type II nanobars in detail [Fig. \ref{Fig4}(j) to \ref{Fig4}(l)] ATA shows a narrower switching field distribution for these nanobars compared to P2A and P3A. About 40\% of Type I and Type II nanobars of ATA undergo reversal within a span of 7.5 mT [Fig. \ref{Fig4}(j)], i.e., within a field regime smaller than for P2A and P3A. Full reversal of the Type I and Type II nanobars is seen at about 34 mT. The corresponding nanobars that are disconnected require a field of up to about 52 mT for full reversal. These observations are qualitatively consistent with the field dependencies of spin wave resonances presented in Fig.~\ref{Fig1b}. We suppose that the narrow field distribution for switching in ATA reflects a cooperative reversal phenomenon like an avalanche [compare the blue chain-like domain in Fig. \ref{Fig4}(d)].\\ \indent The analysis of total charges \textit{Q\textsubscript{tot}} is depicted in Fig. \ref{Fig4}(m) to \ref{Fig4}(o). In all three Ammann tilings we observe a minimum in \textit{Q\textsubscript{tot}} when Type I and Type II nanobars undergo the reversal process. The drop in \textit{Q\textsubscript{tot}} is accompanied by an increased number of FCLs [Fig. \ref{Fig4}(p) to \ref{Fig4}(r)].
In the magnetization reversal of ATB [Fig. \ref{Fig4}(k)] the switching seems to take place in two separate steps [see the increases in the number of reversed nanobars at 28 mT and 42 mT in Fig. \ref{Fig4}(k)]. Consistent with these features, there are local minima (a shoulder and a maximum) in \textit{Q\textsubscript{tot}} (FCLs). We also see that the maximum in FCLs coincides with the minimum in the sum of charges. The slope of reversal in ATC [Fig. \ref{Fig4}(l)] resembles the high-field slope of ATB in Fig. \ref{Fig4}(k) and the ones seen in P2B, P2C and P3C. We argue that the high-field reversal processes which are detected over a broad field regime reflect the disconnected nanobars. In this field regime we find the spin ice rule violations for ATB and ATC [Fig. \ref{Fig4}(t) and \ref{Fig4}(u), respectively] in that the maximum vertex charge \textit{Q} = 3 is present in case of \textit{N} = 3. This large vertex charge is not observed in the reversal of ATA [Fig. \ref{Fig4}(s)].\\ \indent
The data shown in Fig.~\ref{Fig3}(d) are consistent with the evolution of spin-wave branches. Reversal of P3A starts near 30 mT via switching of Type I and Type II nanobars. The switching of Type III nanobars occurs at higher fields. At 52 mT, i.e., the maximum field in the XPEEM microscoype, less than 50 \% have been switched. Consistently, in our spectroscopy data we need to apply an opposing field with $\mu_0|H|=70$~mT to obtain fully developed high-frequency spin-wave branches in Fig.~\ref{Fig1a}(e). In P3B [Fig.~\ref{Fig3}(e)] and P3C [Fig.~\ref{Fig3}(f)] we do not find reversed Type III nanobars up to 52 mT in XPEEM, which explains the faint monotonous spin-wave signals marked by a yellow arrow in Fig.~\ref{Fig1a}(g). The evolution of spin-wave branches measured on ATA [Fig. \ref{Fig1b}(d)] is also consistent with the XPEEM imaging [Fig. \ref{Fig4}(j)]: a large number of Type I and Type II nanobars reverse in a narrow regime between 30 and 34 mT in both experiments. Type III nanobars do not reverse in Fig. \ref{Fig4}(j) explaining the monotonous variation of the branch highlighted by yellow arrows in Fig.~\ref{Fig1b}(d).

\subsection*{Broadband spectroscopy in the minor loop: reconfigurable magnon excitations}
\begin{figure}
	\includegraphics[width=0.5\textwidth]{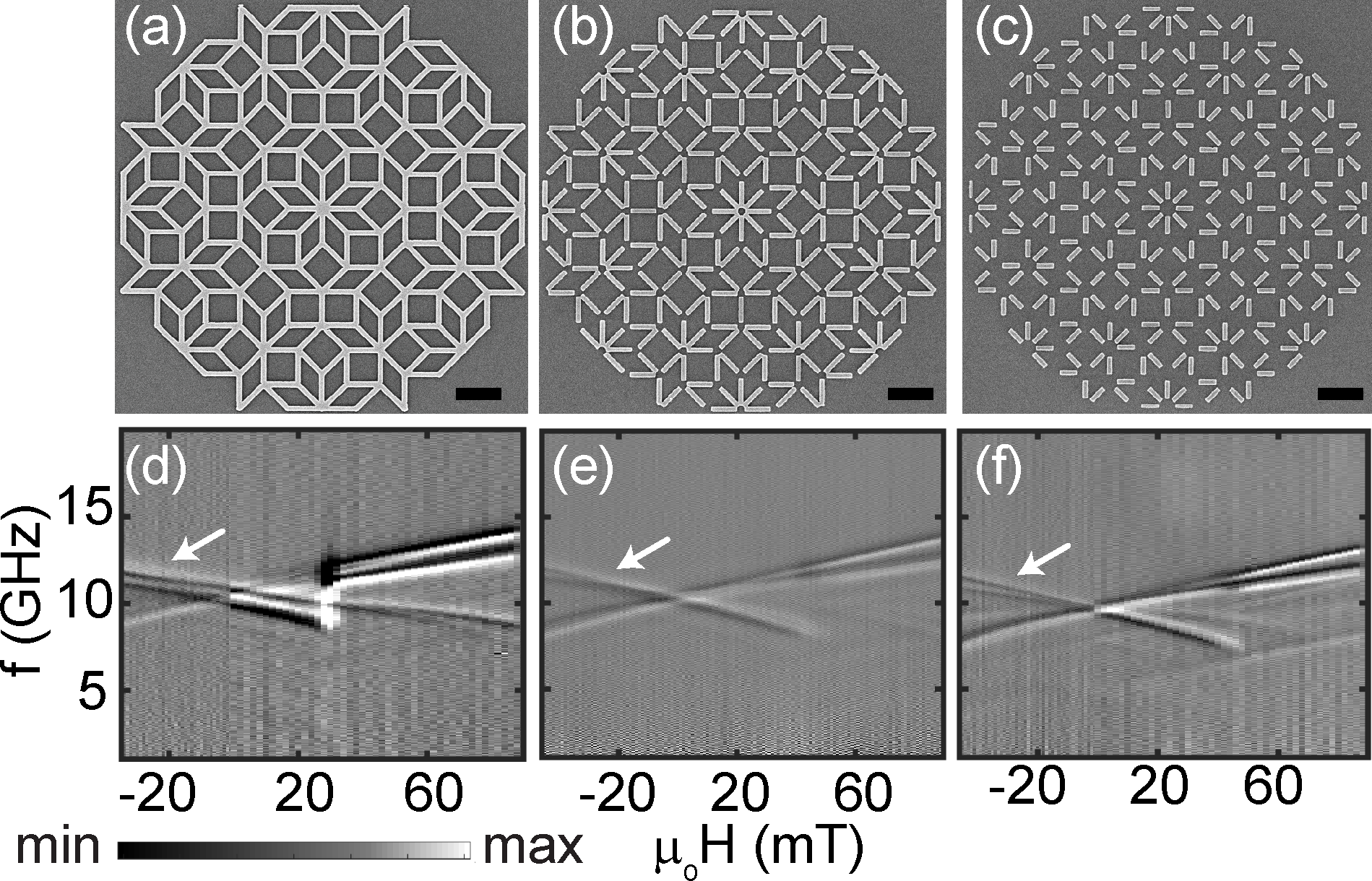}
		\caption{SEM images showing 1st generation Ammann tilings (a) ATA for which nanobars are fully connected, (b) ATB with partially connected nanobars and (c) ATC with disconnected nanobars. The black color scale bar represents 1 micrometer.  Gray-scale spin wave spectra for large arrays  measured on 4th generation  (d)  ATA, (e) ATB, and (f) ATC, respectively.  The minor loop magnetic field protocol implemented was (d) +90 mT $ \rightarrow $ -34 mT	$ \rightarrow $ 90 mT, (e) +90 mT $ \rightarrow $ -46 mT	$ \rightarrow $ 90 mT, and (f) +90 mT $ \rightarrow $ -46 mT	$ \rightarrow $ 90 mT, respectively. }\label{Fig6}
\end{figure}

In the following we discuss the magnonic excitations which are detected in minor loop measurements starting from intermediate negative fields after magnetizing ATA, ATB and ATC at $+90~$mT. In Fig.~\ref{Fig6} we show color-coded spin-wave spectra taken on partially reversed quasicrystals. In each case the depicted spectra belong to a minor loop which starts near the magnetic field value for which the XPEEM data suggest a minimum in the sum of vertex charges and a large number of flux-closure loops for the Ammann tilings. XPEEM images for ATA, ATB, and ATC were shown Fig.~\ref{Fig3}(d) - (f), respectively. For these three samples the branches in the minor loop [Fig.~\ref{Fig6}(d) - (f)]  are more complex compared to major loop measurements. They exhibit both $df/dH>0$ and $df/dH<0$ for $H<0$ reflecting non-reversed and reversed nanobars, respectively. The white arrows highlight branches which were not observed in Fig. \ref{Fig1b}(d) - (f). The number of flux-closure loops does not seem to be large enough to resolve additional x-shaped magnon branches near $H=0$ which are characteristic for spin dynamics in magnetic vortex configurations \cite{Giesen2005,pod2006}. All the reprogrammed branches in Fig.~\ref{Fig6} exhibit small linewidths which are promising in view of magnonic functionalities like reprogrammable frequency filters.

\section*{Conclusions}
To conclude, we varied the exchange and dipolar interactions in Penrose P2, P3, and Ammann quasicrystal lattices. We explored magnonic excitations and imaged ferromagnetic reversal in these exotic ASI structures using XPEEM. Our data show compliance with ice rules in the exchange coupled nanobars. Ice rule violation occurs prominently in tilings without exchange coupling among nanobars. Owing to the asymmetric and aperiodic arrangements around each vertex, nanobars taking the same angle with the applied field have significantly different switching behavior depending on their local environment. We observe the narrowest distribution of reversal fields in an exchange-coupled Ammann tiling, followed by the exchange-coupled Penrose P2 and P3 tilings. Here reversal is triggered by vertices whose charge \textit{Q} deviates from the ground state. We demonstrated that spin-wave resonances remain sharp for the partially reversed quasicrystals. The exchange-coupled Penrose P3 and Ammann tilings show the formation of extended domains via non-stochastic cooperative reversal which might be functionalized for reprogrammable magnon steering.

\section*{Methods}
\subsection*{Sample Fabrication} A bilayer PMMA/MMA resist was spin-coated on a silicon substrate, and exposed via Raith electron beam lithography system at 100 KeV. After development of the resist a 25 nm thick Py film was deposited using ebeam evaporator. Subsequently, ultrasonic assisted lift-off was performed in N-methyl Pyrilidone solution.
\subsection*{XPEEM Measurements} XPEEM imaging was done at the SPEEM station located at the UE49/PGMa beamline at BESSY-II (Helmholtz Zentrum, Berlin). The samples were mounted on a sample holder which allowed us to apply magnetic field to the sample in-situ. The Penrose P2 and P3 AMQs were patterned on the same silicon substrate, whereas Ammann AMQs were fabricated on another identical silicon substrate. The Magnetic images were obtained by performing XMCD at the Fe L3-edge. The obtained contrast is a measure of the projection of the magnetization on the X-ray polarization vector. Thus nanobars with a magnetization parallel or antiparallel to the X-ray polarization either appear red or blue.  The sample orientation was optimized to maximize the contrast.
\subsection*{Broadband spin-wave spectroscopy}
 To detect the resonances, we connected two ports of a vector network analyzer to both ends of the CPW using microwave probes and coaxial cables. Subsequently, we applied a constant global magnetic field (from $\pm$ 90 mT  to $ \mp $90  mT  in steps of $\pm$1 mT) at a given in-plane angle $\phi$ and performed frequency sweeps from 1 GHz up to 20 GHz using a vector network analyzer. We collect the S-parameters from the vector network analyzer as a function of frequency at the constant applied magnetic field, $\mu_{o}H$, and angle $ \phi $, and this corresponds to single spin-wave spectra at one $\mu_{o}H$ and $ \phi $ value. We then subtracted spin-wave spectra taken at successive fields $\mu_{o}H$; that is, we obtain $ \Delta $ S (i) = S [H (i+1), $ \phi $] - S [H (i), $ \phi $].  
\subsection*{Micromagnetic Simulations} Simulations were performed using the OOMMF code \cite{OOMMF1}, and the Py parameters used in simulations were as follows: Exchange constant \textit{A} = 1.3 $ \times $ 10 \textsuperscript{-11} J m\textsuperscript{-1}, saturation magnetization \textit{M\textsubscript{S}} = 8 $ \times $  10\textsuperscript{5}  A m\textsuperscript{-1}, magnetocrystalline anisotropy constant \textit{K} = 0, gyromagnetic ratio $ \gamma $  = 2.211 $ \times $  10\textsuperscript{5} m A\textsuperscript{-1} s\textsuperscript{-1}, and dimensionless damping coefficient $ \alpha $  = 0.01. Different magnetic configurations were explored by initializing magnetization vectors of individual segments and relaxing the spin system at the given magnetic field. For this, we first created a colored bitmap (on a grid of 5 nm$ \times $ 5 nm $ \times $ 25 nm) where each segment was assigned a color corresponding to its assumed magnetization orientation. We then imported this colored bitmap into OOMMF and equilibrated it in the presence of a field that resided within the experimental switching regime. \\

\subsection*{Charge model}
The charge model (CM) assigns a magnetic charge to each of the vertices for evaluating the magnetic energy \cite{castelnovo2008magnetic}. For this one assumes each nanobar to be a dumbbell of length \textit{l} with two equal charges of opposite polarity, $\pm $\textit{q} = $\pm $ \textit{m}/\textit{l} = $\pm $ \textit{Mtwl}/\textit{l} = $\pm $ \textit{Mtw}, at the dumbbell’s ends. \textit{m}, \textit{M}, \textit{t}, and \textit{w} represent the magnetic dipole moment, saturation magnetization, thickness, and width of a segment, respectively. A vertex with a coordination number (CN) \textit{N} can acquire a charge \textit{Q} = $ \sum $ \textit{q} = +\textit{Nq} $ \dots $ -\textit{Nq} which is the sum of individual charges \textit{q}. For a given CN a vertex acquires the lowest possible charge \textit{Q} in the ground state to minimize the total energy. 

\section*{Data Availability}
Requests concerning data should be addressed to D.G.
The datasets analysed in the current study are available in the Zenodo repository, ....

\section*{Code Availability}
The code used for micromagnetic simulations is found in Ref.~\cite{OOMMF1} and input scripts are available in the Zenodo repository, ....

\bibliography{Bibliography_1}

\begin{thebibliography}{10}
\expandafter\ifx\csname url\endcsname\relax
  \def\url#1{\texttt{#1}}\fi
\expandafter\ifx\csname urlprefix\endcsname\relax\def\urlprefix{URL }\fi
\providecommand{\bibinfo}[2]{#2}
\providecommand{\eprint}[2][]{\url{#2}}

\bibitem{shechtman1984metallic}
\bibinfo{author}{Shechtman, D.}, \bibinfo{author}{Blech, I.},
  \bibinfo{author}{Gratias, D.} \& \bibinfo{author}{Cahn, J.~W.}
\newblock \bibinfo{title}{Metallic phase with long-range orientational order
  and no translational symmetry}.
\newblock \emph{\bibinfo{journal}{Phys. Rev. Lett.}}
  \textbf{\bibinfo{volume}{53}}, \bibinfo{pages}{1951} (\bibinfo{year}{1984}).

\bibitem{goldman2013family}
\bibinfo{author}{Goldman, A.~I.} \emph{et~al.}
\newblock \bibinfo{title}{A family of binary magnetic icosahedral quasicrystals
  based on rare earths and cadmium}.
\newblock \emph{\bibinfo{journal}{Nat. Mater.}} \textbf{\bibinfo{volume}{12}},
  \bibinfo{pages}{714--718} (\bibinfo{year}{2013}).

\bibitem{bhat2013controlled}
\bibinfo{author}{Bhat, V.~S.} \emph{et~al.}
\newblock \bibinfo{title}{Controlled magnetic reversal in permalloy films
  patterned into artificial quasicrystals}.
\newblock \emph{\bibinfo{journal}{Phys. Rev. Lett.}}
  \textbf{\bibinfo{volume}{111}}, \bibinfo{pages}{077201}
  (\bibinfo{year}{2013}).

\bibitem{wang2006artificial}
\bibinfo{author}{Wang, R.} \emph{et~al.}
\newblock \bibinfo{title}{Artificial ‘spin ice’in a geometrically
  frustrated lattice of nanoscale ferromagnetic islands}.
\newblock \emph{\bibinfo{journal}{Nature}} \textbf{\bibinfo{volume}{439}},
  \bibinfo{pages}{303} (\bibinfo{year}{2006}).

\bibitem{qi2008direct}
\bibinfo{author}{Qi, Y.}, \bibinfo{author}{Brintlinger, T.} \&
  \bibinfo{author}{Cumings, J.}
\newblock \bibinfo{title}{Direct observation of the ice rule in an artificial
  kagome spin ice}.
\newblock \emph{\bibinfo{journal}{Phys. Rev. B}} \textbf{\bibinfo{volume}{77}},
  \bibinfo{pages}{094418} (\bibinfo{year}{2008}).

\bibitem{mengotti2011real}
\bibinfo{author}{Mengotti, E.} \emph{et~al.}
\newblock \bibinfo{title}{Real-space observation of emergent magnetic monopoles
  and associated dirac strings in artificial kagome spin ice}.
\newblock \emph{\bibinfo{journal}{Nat. Phys.}} \textbf{\bibinfo{volume}{7}},
  \bibinfo{pages}{68--74} (\bibinfo{year}{2011}).

\bibitem{castelnovo2008magnetic}
\bibinfo{author}{Castelnovo, C.}, \bibinfo{author}{Moessner, R.} \&
  \bibinfo{author}{Sondhi, S.~L.}
\newblock \bibinfo{title}{Magnetic monopoles in spin ice}.
\newblock \emph{\bibinfo{journal}{Nature}} \textbf{\bibinfo{volume}{451}},
  \bibinfo{pages}{42--45} (\bibinfo{year}{2008}).

\bibitem{branford2012emerging}
\bibinfo{author}{Branford, W.}, \bibinfo{author}{Ladak, S.},
  \bibinfo{author}{Read, D.}, \bibinfo{author}{Zeissler, K.} \&
  \bibinfo{author}{Cohen, L.}
\newblock \bibinfo{title}{Emerging chirality in artificial spin ice}.
\newblock \emph{\bibinfo{journal}{Science}} \textbf{\bibinfo{volume}{335}},
  \bibinfo{pages}{1597--1600} (\bibinfo{year}{2012}).

\bibitem{Keswani2019}
\bibinfo{author}{Keswani, N.} \& \bibinfo{author}{Das, P.}
\newblock \bibinfo{title}{On the micromagnetic behavior of dipolar-coupled
  nanomagnets in defective square artificial spin ice systems}.
\newblock \emph{\bibinfo{journal}{J. Appl. Phys.}}
  \textbf{\bibinfo{volume}{126}}, \bibinfo{pages}{214304}
  (\bibinfo{year}{2019}).

\bibitem{AdvASI2020}
\bibinfo{author}{Skjærvø, S.~H.}, \bibinfo{author}{Marrows, C.~H.},
  \bibinfo{author}{Stamps, R.~L.} \& \bibinfo{author}{Heyderman, L.~J.}
\newblock \bibinfo{title}{Advances in artificial spin ice}.
\newblock \emph{\bibinfo{journal}{Nat. Rev. Phys.}}
  \textbf{\bibinfo{volume}{2}}, \bibinfo{pages}{13--28} (\bibinfo{year}{2020}).

\bibitem{gilbert2015PRB}
\bibinfo{author}{Gilbert, I.} \emph{et~al.}
\newblock \bibinfo{title}{Direct visualization of memory effects in artificial
  spin ice}.
\newblock \emph{\bibinfo{journal}{Phys. Rev. B}} \textbf{\bibinfo{volume}{92}},
  \bibinfo{pages}{104417} (\bibinfo{year}{2015}).

\bibitem{PhysRevB.100.214425}
\bibinfo{author}{Arroo, D.~M.}, \bibinfo{author}{Gartside, J.~C.} \&
  \bibinfo{author}{Branford, W.~R.}
\newblock \bibinfo{title}{Sculpting the spin-wave response of artificial spin
  ice via microstate selection}.
\newblock \emph{\bibinfo{journal}{Phys. Rev. B}}
  \textbf{\bibinfo{volume}{100}}, \bibinfo{pages}{214425}
  (\bibinfo{year}{2019}).

\bibitem{gliga2013spectral}
\bibinfo{author}{Gliga, S.}, \bibinfo{author}{K{\'a}kay, A.},
  \bibinfo{author}{Hertel, R.} \& \bibinfo{author}{Heinonen, O.~G.}
\newblock \bibinfo{title}{Spectral analysis of topological defects in an
  artificial spin-ice lattice}.
\newblock \emph{\bibinfo{journal}{Phys. Rev. Lett.}}
  \textbf{\bibinfo{volume}{110}}, \bibinfo{pages}{117205}
  (\bibinfo{year}{2013}).

\bibitem{PhysRevB.93.134420}
\bibinfo{author}{Iacocca, E.}, \bibinfo{author}{Gliga, S.},
  \bibinfo{author}{Stamps, R.~L.} \& \bibinfo{author}{Heinonen, O.}
\newblock \bibinfo{title}{Reconfigurable wave band structure of an artificial
  square ice}.
\newblock \emph{\bibinfo{journal}{Phys. Rev. B}} \textbf{\bibinfo{volume}{93}},
  \bibinfo{pages}{134420} (\bibinfo{year}{2016}).

\bibitem{jungfleisch2016dynamic}
\bibinfo{author}{Jungfleisch, M.} \emph{et~al.}
\newblock \bibinfo{title}{Dynamic response of an artificial square spin ice}.
\newblock \emph{\bibinfo{journal}{Phys. Rev. B}} \textbf{\bibinfo{volume}{93}},
  \bibinfo{pages}{100401} (\bibinfo{year}{2016}).

\bibitem{doi:10.1063/1.4978315}
\bibinfo{author}{Li, Y.} \emph{et~al.}
\newblock \bibinfo{title}{Thickness dependence of spin wave excitations in an
  artificial square spin ice-like geometry}.
\newblock \emph{\bibinfo{journal}{J. Appl. Phys.}}
  \textbf{\bibinfo{volume}{121}}, \bibinfo{pages}{103903}
  (\bibinfo{year}{2017}).

\bibitem{Lendinez_2019}
\bibinfo{author}{Lendinez, S.} \& \bibinfo{author}{Jungfleisch, M.~B.}
\newblock \bibinfo{title}{Magnetization dynamics in artificial spin ice}.
\newblock \emph{\bibinfo{journal}{J. Phys.: Condens. Matter}}
  \textbf{\bibinfo{volume}{32}}, \bibinfo{pages}{013001}
  (\bibinfo{year}{2019}).

\bibitem{PhysRevB.100.054433}
\bibinfo{author}{Dion, T.} \emph{et~al.}
\newblock \bibinfo{title}{Tunable magnetization dynamics in artificial spin ice
  via shape anisotropy modification}.
\newblock \emph{\bibinfo{journal}{Phys. Rev. B}}
  \textbf{\bibinfo{volume}{100}}, \bibinfo{pages}{054433}
  (\bibinfo{year}{2019}).

\bibitem{iacocca2020tailoring}
\bibinfo{author}{Iacocca, E.}, \bibinfo{author}{Gliga, S.} \&
  \bibinfo{author}{Heinonen, O.~G.}
\newblock \bibinfo{title}{Tailoring spin-wave channels in a reconfigurable
  artificial spin ice}.
\newblock \emph{\bibinfo{journal}{Physical Review Applied}}
  \textbf{\bibinfo{volume}{13}}, \bibinfo{pages}{044047}
  (\bibinfo{year}{2020}).

\bibitem{Li_2016}
\bibinfo{author}{Li, Y.} \emph{et~al.}
\newblock \bibinfo{title}{Brillouin light scattering study of magnetic-element
  normal modes in a square artificial spin ice geometry}.
\newblock \emph{\bibinfo{journal}{J. Phys. D: Appl. Phys.}}
  \textbf{\bibinfo{volume}{50}}, \bibinfo{pages}{015003}
  (\bibinfo{year}{2016}).

\bibitem{brajuskovic2016real}
\bibinfo{author}{Brajuskovic, V.}, \bibinfo{author}{Barrows, F.},
  \bibinfo{author}{Phatak, C.} \& \bibinfo{author}{Petford-Long, A.}
\newblock \bibinfo{title}{Real space observation of magnetic excitations and
  avalanche behavior in artificial quasicrystal lattices}.
\newblock \emph{\bibinfo{journal}{Sci. Rep.}} \textbf{\bibinfo{volume}{6}},
  \bibinfo{pages}{34384} (\bibinfo{year}{2016}).

\bibitem{farmer2016direct}
\bibinfo{author}{Farmer, B.} \emph{et~al.}
\newblock \bibinfo{title}{Direct imaging of coexisting ordered and frustrated
  sublattices in artificial ferromagnetic quasicrystals}.
\newblock \emph{\bibinfo{journal}{Phys. Rev. B}} \textbf{\bibinfo{volume}{93}},
  \bibinfo{pages}{134428} (\bibinfo{year}{2016}).

\bibitem{bhat2014non}
\bibinfo{author}{Bhat, V.} \emph{et~al.}
\newblock \bibinfo{title}{Non-stochastic switching and emergence of magnetic
  vortices in artificial quasicrystal spin ice}.
\newblock \emph{\bibinfo{journal}{Physica C: Superconductivity and its
  Applications}} \textbf{\bibinfo{volume}{503}}, \bibinfo{pages}{170--174}
  (\bibinfo{year}{2014}).

\bibitem{shi2018frustration}
\bibinfo{author}{Shi, D.} \emph{et~al.}
\newblock \bibinfo{title}{Frustration and thermalization in an artificial
  magnetic quasicrystal}.
\newblock \emph{\bibinfo{journal}{Nat. Phys.}} \textbf{\bibinfo{volume}{14}},
  \bibinfo{pages}{309} (\bibinfo{year}{2018}).

\bibitem{WangSci2016}
\bibinfo{author}{Wang, Y.-L.} \emph{et~al.}
\newblock \bibinfo{title}{Rewritable artificial magnetic charge ice}.
\newblock \emph{\bibinfo{journal}{Science}} \textbf{\bibinfo{volume}{352}},
  \bibinfo{pages}{962} (\bibinfo{year}{2016}).

\bibitem{Demo2017}
\bibinfo{author}{Stamps, R.~L.} \emph{et~al.}
\newblock \emph{\bibinfo{title}{Spin Waves On Spin Structures: Topology,
  Localization, And Nonreciprocity}}, chap.~\bibinfo{chapter}{8},
  \bibinfo{pages}{219--260} (\bibinfo{publisher}{CRC Press},
  \bibinfo{year}{2017}), \bibinfo{edition}{2} edn.

\bibitem{penrose1979math}
\bibinfo{author}{Penrose, R.}
\newblock \bibinfo{title}{Pentaplexity - a class of non-periodic tilings of the
  plane}.
\newblock \emph{\bibinfo{journal}{The Mathematical Intelligencer}}
  \textbf{\bibinfo{volume}{2}}, \bibinfo{pages}{32} (\bibinfo{year}{1979}).

\bibitem{Sethna1997}
\bibinfo{author}{Perković, O.} \& \bibinfo{author}{Sethna, J.~P.}
\newblock \bibinfo{title}{Improved magnetic information storage using
  return-point memory}.
\newblock \emph{\bibinfo{journal}{J. Appl. Phys.}}
  \textbf{\bibinfo{volume}{81}}, \bibinfo{pages}{1590--1597}
  (\bibinfo{year}{1997}).

\bibitem{aharoni1998demagnetizing}
\bibinfo{author}{Aharoni, A.}
\newblock \bibinfo{title}{Demagnetizing factors for rectangular ferromagnetic
  prisms}.
\newblock \emph{\bibinfo{journal}{J. Appl. Phys.}}
  \textbf{\bibinfo{volume}{83}}, \bibinfo{pages}{3432--3434}
  (\bibinfo{year}{1998}).

\bibitem{burn2015angular}
\bibinfo{author}{Burn, D.}, \bibinfo{author}{Chadha, M.} \&
  \bibinfo{author}{Branford, W.}
\newblock \bibinfo{title}{Angular-dependent magnetization reversal processes in
  artificial spin ice}.
\newblock \emph{\bibinfo{journal}{Phys. Rev. B}} \textbf{\bibinfo{volume}{92}},
  \bibinfo{pages}{214425} (\bibinfo{year}{2015}).

\bibitem{Bhat2018}
\bibinfo{author}{Bhat, V.~S.} \& \bibinfo{author}{Grundler, D.}
\newblock \bibinfo{title}{Angle-dependent magnetization dynamics with
  mirror-symmetric excitations in artificial quasicrystalline nanomagnet
  lattices}.
\newblock \emph{\bibinfo{journal}{Phys. Rev. B}} \textbf{\bibinfo{volume}{98}},
  \bibinfo{pages}{174408} (\bibinfo{year}{2018}).

\bibitem{Gurevich96}
\bibinfo{author}{Gurevich, A.~G.} \& \bibinfo{author}{Melkov, G.~A.}
\newblock \emph{\bibinfo{title}{Magnetization oscillations and waves}}
  (\bibinfo{publisher}{CRC Press}, \bibinfo{year}{1996.}).

\bibitem{Giesen2005}
\bibinfo{author}{Giesen, F.} \emph{et~al.}
\newblock \bibinfo{title}{Hysteresis and control of ferromagnetic resonances in
  rings}.
\newblock \emph{\bibinfo{journal}{Appl. Phys. Lett.}}
  \textbf{\bibinfo{volume}{86}}, \bibinfo{pages}{112510}
  (\bibinfo{year}{2005}).

\bibitem{pod2006}
\bibinfo{author}{Podbielski, J.}, \bibinfo{author}{Giesen, F.} \&
  \bibinfo{author}{Grundler, D.}
\newblock \bibinfo{title}{Spin-wave interference in microscopic rings}.
\newblock \emph{\bibinfo{journal}{Phys. Rev. Lett.}}
  \textbf{\bibinfo{volume}{96}}, \bibinfo{pages}{167207}
  (\bibinfo{year}{2006}).

\bibitem{OOMMF1}
\bibinfo{author}{Donahue, M.} \& \bibinfo{author}{Porter, D.}
\newblock \bibinfo{title}{Oommf user guide, version 1.0}.
\newblock \emph{\bibinfo{journal}{Interagency Report NISTIR 6376}}
  (\bibinfo{year}{1999}).

\end{thebibliography}
\bibliographystyle{naturemag}

\section*{Acknowledgements}
The research was supported by the Swiss National Science Foundation via Grant No. 163016. We thank HZB for the allocation of synchrotron radiation beamtime. V.S. Bhat acknowledges support from the foundation for Polish Science through the IRA Programme financed by EU within SG OP Programme.

\section*{Author contributions statement}
V.B. and D.G. conceived the project. V.B. fabricated the samples and performed the micromagnetic simulations.  V.B., S.W., F.K, and K.B. conducted the XPEEM experiments. V.B., S.W., and D.G performed the data analysis and wrote the manuscript on which all authors commented.

\section*{Competing interests}
The authors declare no competing interests.

\section*{Materials \& Correspondence} Correspondence and material requests should be addressed to D.G.

\end{document}